\documentclass[aps,prx,reprint,amsmath,amssymb,superscriptaddress,showpacs,nobibnotes]{revtex4-1}

\usepackage{graphicx}
\usepackage{bm}     
\usepackage{hyperref} 
\usepackage{dsfont}    
\usepackage{color}
\usepackage[normalem]{ulem}
\usepackage{cleveref}

\newcommand{\ket}[1]{|#1\rangle}

\newcommand{\tr}{\text{tr}}
\newcommand{\abs}[1]{\left|{#1}\right|}

\newcommand{\eq}[1]{Eq.~(\ref{#1})}
\newcommand{\fig}[1]{Fig.~\ref{#1}}

\def\beq{\begin{eqnarray}}
\def\eeq{\end{eqnarray}}

\def\B{\mathcal{B}}

\def\S{\mathcal{S}}
\def\R{\mathcal{R}}
\def\L{\mathcal{L}}

\newcommand\redout{\bgroup\markoverwith{\textcolor{red}{\rule[.5ex]{2pt}{1.2pt}}}\ULon}

\begin{document}

\title{Tomography of a Mode-Tunable Coherent Single-Photon Subtractor}

\author{Young-Sik Ra}
\affiliation{Laboratoire Kastler Brossel, UPMC-Sorbonne Universit\'es, CNRS, ENS-PSL Research University, Coll\`{e}ge de France; 4 place Jussieu, 75252 Paris, France}

\author{Cl\'ement Jacquard}
\affiliation{Laboratoire Kastler Brossel, UPMC-Sorbonne Universit\'es, CNRS, ENS-PSL Research University, Coll\`{e}ge de France; 4 place Jussieu, 75252 Paris, France}

\author{Adrien Dufour}
\affiliation{Laboratoire Kastler Brossel, UPMC-Sorbonne Universit\'es, CNRS, ENS-PSL Research University, Coll\`{e}ge de France; 4 place Jussieu, 75252 Paris, France}

\author{Claude Fabre}
\affiliation{Laboratoire Kastler Brossel, UPMC-Sorbonne Universit\'es, CNRS, ENS-PSL Research University, Coll\`{e}ge de France; 4 place Jussieu, 75252 Paris, France}

\author{Nicolas Treps}
\affiliation{Laboratoire Kastler Brossel, UPMC-Sorbonne Universit\'es, CNRS, ENS-PSL Research University, Coll\`{e}ge de France; 4 place Jussieu, 75252 Paris, France}

\date{\today}

\begin{abstract} 
Single-photon subtraction plays important roles in optical quantum information processing as it provides a non-Gaussian characteristic in continuous-variable quantum information. While the conventional way of implementing single-photon subtraction based on a low-reflectance beam splitter works properly for a single-mode quantum state, it is unsuitable for a multimode quantum state because a single photon is subtracted from all multiple modes without maintaining their mode coherence. Here we experimentally implement and characterize a mode-tunable coherent single-photon subtractor based on sum frequency generation. It can subtract a single photon exclusively from one desired time-frequency mode of light or from a coherent superposition of multiple time-frequency modes. To fully characterize the implemented single-photon subtractions, we employ quantum process tomography based on coherent states. The mode-tunable coherent single-photon subtractor will be an essential element for realizing non-Gaussian quantum networks necessary to get a quantum advantage in information processing.

\end{abstract}

\pacs{
42.50.Ex, 
03.65.Wj,	
42.65.Ky	
}

\maketitle


\section{Introduction}

Optical quantum information processing can be classified mainly into two approaches depending on encoding of quantum information: one is based on continuous electric field quadratures (thus, referred to as continuous-variable quantum information), and the other is based on discrete photon numbers (discrete-variable quantum information). Each of the approaches has its own advantages compared with the other: e.g., in the continuous-variable approach, highly multimode entangled states can be deterministically generated~\cite{Weedbrook:2012fe,Yokoyama:2013jp,Roslund:2014cb,Chen:2014jx,Cai:2016we}, and in the discrete-variable approach, quantum processes that cannot be classically simulated can be implemented~\cite{Broome:2013bv,Spring:2013do,Crespi:2013fu,Tillmann:2013jv}. Therefore, a new approach to combine both advantages has attracted much attention, which is called hybrid quantum information processing~\cite{Andersen:2015dp}. One of the fundamental operations for the hybrid approach is single-photon subtraction, mathematically described by the annihilation operator $\hat{a}$. It introduces a non-Gaussian characteristic (i.e. negativity of Wigner function) in continuous-variable quantum information~\cite{Wenger:2004cw}, which plays essential roles in various quantum information processing, e.g., universal~\cite{Lloyd:1999vz,Menicucci:2006ir} and genuine~\cite{Bartlett:2002fo,Mari:2012ep} quantum computing, preparation of coherent-state-superposition~\cite{Ourjoumtsev:2006jn,NeergaardNielsen:2006hl,Ourjoumtsev:2009jh,NeergaardNielsen:2010by} and hybrid entanglement~\cite{Jeong:2014bl,Morin:2014ip}, noiseless linear amplification~\cite{Zavatta:2011ea}, and entanglement concentration~\cite{Ourjoumtsev:2007he,Takahashi:2010kw}.

The conventional way of implementing the single-photon subtraction is to detect a single photon tapped from an input light using a low-reflectance beam splitter~\cite{Wenger:2004cw,Kim:2008bw}. Such a method works well for a single-mode state~\cite{Ourjoumtsev:2006jn,NeergaardNielsen:2006hl,NeergaardNielsen:2010by}, but it is unsuitable for a multimode state because the detected photon comes from any mode in an incoherent way, which results in a complete mixture of annihilation operators over the multiple modes~\cite{Averchenko:2016gv}. To fully benefit from the highly multimode entangled states available in the continuous variable approach~\cite{Weedbrook:2012fe,Yokoyama:2013jp,Roslund:2014cb,Chen:2014jx,Cai:2016we}, one accordingly requires a single-photon subtraction that is able to operate only in the desired modes by maintaining their mode coherence~\cite{Ourjoumtsev:2009jh,Ourjoumtsev:2007he,Kim:2013bt}.

In this work, we implement and characterize a single-photon subtractor which can be tuned to subtract a single photon exclusively from one desired time-frequency mode of light or coherently from multiple time-frequency modes. The single-photon subtractor is based on the detection of a single photon generated via a sum frequency interaction between an input beam and a strong gate beam in which the choice of the gate-beam modes determines the time-frequency modes of single-photon subtraction~\cite{Eckstein:2011jp,Brecht:2014eg,Averchenko:2014dn,Manurkar:2016hx}. To characterize single-photon subtractions with various choices of the gate beam modes, we measure the \emph{subtraction matrix} of each single-photon subtraction by employing coherent-state quantum process tomography~\cite{Lobino:2008p6659,Fedorov:2015gg}: the subtraction matrix contains complete information about a general single-photon subtraction (i.e. amplitude, phase, and coherence between different modes), and can be used to quantify its performances. We furthermore discuss the possible experimental imperfections in a single-photon subtractor such as unwanted heralding events (e.g., dark counts or two-photon detection) and optical losses, and estimate their effect on preparing a non-Gaussian quantum state.

\section{Description of a general single-photon subtraction}

\begin{figure}[t]
\centerline{\includegraphics[width=0.45\textwidth]{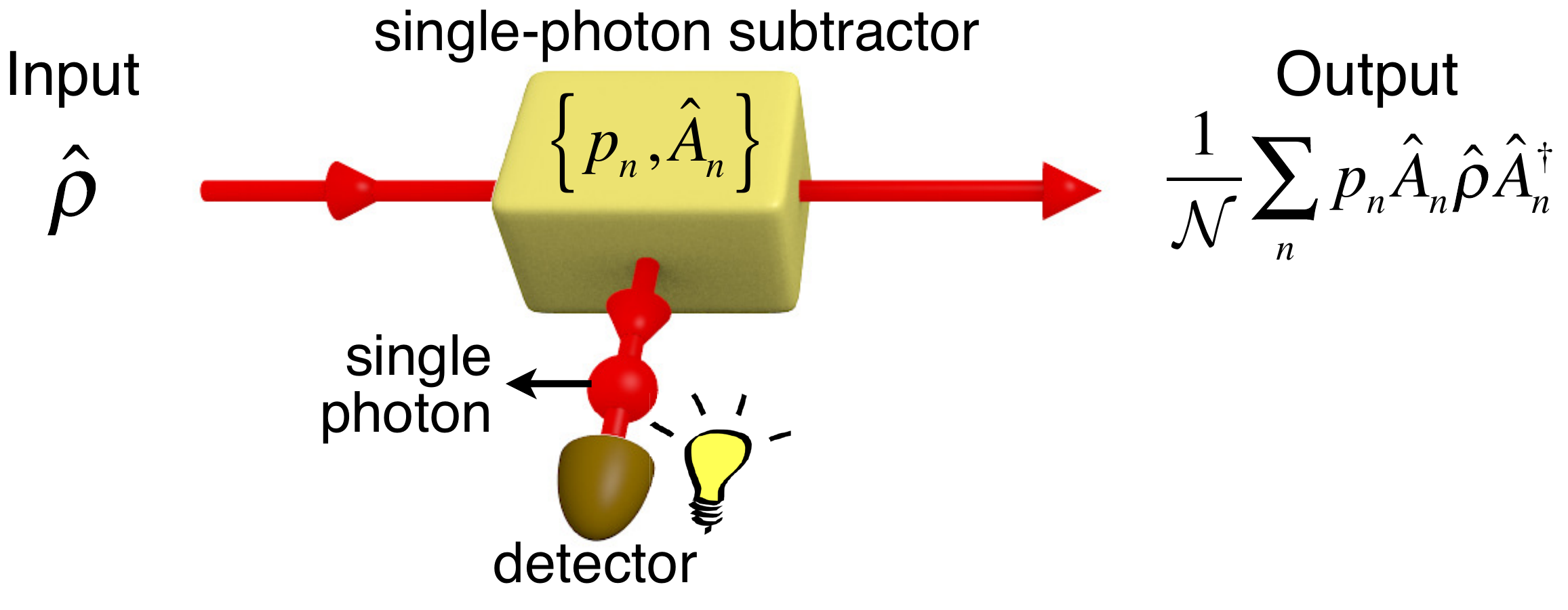}}
\caption{A single-photon subtractor removes exactly one photon from an input state $\hat{\rho}$, which is heralded by the detection of a single photon in the ancillary path. Single-photon subtraction can be described, in general, by a mixture of annihilation operators $\hat{A}_0$, $\hat{A}_1$, $\dots$ with the corresponding weights $p_0$, $p_1$, $\dots$, where all weights sum to one, and different annihilation operators are not necessarily orthogonal, $[ \hat{A}_n, \hat{A}^\dagger_m]\ne 0$. The normalization constant $\mathcal{N}$ is $\sum_n p_n \langle \hat{A}^\dagger_n \hat{A}_n \rangle $, which is proportional to the heralding probability.
}\label{fig:description}
\end{figure}

We start by introducing a formalism describing a general single-photon subtraction in multiple modes~\cite{Averchenko:2016gv}. In a single-mode case, single-photon subtraction is uniquely defined by the single-photon annihilation operator $\hat{a}$, which lowers one excitation of a photon-number state $\ket{n}$: $\hat{a}\ket{n}=\sqrt{n}\ket{n-1}$. This operation is intrinsically nondeterministic (i.e. non-trace-preserving)~\cite{Kumar:2013ic}, which succeeds only if a desired outcome is obtained as measuring an ancillary system~\cite{Wenger:2004cw,Kim:2008bw}. In the multimode case, on the other hand, single-photon subtraction can be diverse because it can consist of, for example, one annihilation operator from multiple modes or several annihilation operators from multiple modes, added as a superposition or as a mixture. In general, single-photon subtraction can be described by a mixture of annihilation operators $\hat{A}_n$ with weights $p_n$, as shown in \fig{fig:description}, where $\hat{A}_n$ can be expressed as a linear combination of basis annihilation operators $\{\hat{a}_0, \hat{a}_1, \dots, \hat{a}_{d-1} \}$ in a $d$-dimensional orthonormal mode basis: $\hat{A}_n=\sum_{i=0} ^{d-1} c_{ni} \hat{a}_i$. The bosonic commutation relation of each annihilation operator $[ \hat{A}_n, \hat{A}^\dagger_n ] = 1 $ dictates that $\sum_i \abs{c_{ni}}^2 =1$, but different annihilation operators are not necessarily orthogonal, $[ \hat{A}_n, \hat{A}^\dagger_m]\ne 0$. A single-photon subtraction $\S$ acting on an input state $\hat{\rho}$ can then be expressed as a quantum map
\beq
\S[\hat{\rho}] = \sum_n p_n \hat{A}_n \hat{\rho} \hat{A}_n^\dagger = \sum_{i,j=0}^{d-1} \chi_{ij} \hat{a}_i \hat{\rho} \hat{a}^\dagger_j,
\label{eq:subtractor}
\eeq
where $\chi_{ij}=\sum_n p_n c_{ni} c_{nj}^*$. It results in the output state $\S[\hat{\rho}]/\tr{(\S[\hat{\rho}])}$ with success probability proportional to $\tr{(\S[\hat{\rho}])}=\sum_{ij}^{d-1} \chi_{ij} \langle \hat{a}^\dagger_j \hat{a}_i \rangle $. This formalism can also be obtained from single-photon subtraction based on a multimode beamsplitter as reported in Ref.~\cite{Averchenko:2016gv}. It is important to note that a single-photon subtraction $\S$ is uniquely determined by the \emph{subtraction matrix} $\chi$, which is analogous to the density matrix representation for a quantum state. The subtraction matrix is Hermitian and positive semidefinite with trace of one, and $\tr(\chi^2)$ quantifies the purity of the operation, $1/\tr(\chi^2)$ the effective number of orthogonal modes, and $(\tr\sqrt{ \sqrt{\chi} \mu \sqrt{\chi} })^2$ the fidelity between two single-photon subtractions described by $\chi$ and $\mu$.

As an example, a single-photon subtractor based on the conventional method~\cite{Wenger:2004cw,Kim:2008bw} makes a completely incoherent single-photon subtraction
$\S^{\text{(incoh)}}[\hat{\rho}] = \sum_{i=0}^{d-1} \frac{1}{d} \hat{a}_i \hat{\rho} \hat{a}_i^\dagger,$
which gives rise to the identity subtraction matrix $\chi^{\text{(incoh)}}_{ij}$=$\delta_{ij}/d$ exhibiting purity of $1/d$.
On the other hand, a coherent single-photon subtraction
$\S^{\text{(coh)}}[\hat{\rho}] = \hat{A}_0 \hat{\rho} \hat{A}_0^\dagger$
with $\hat{A}_0=\sum_{i=0}^{d-1} c_i \hat{a}_i$ shows the subtraction matrix of $\chi^{\text{(coh)}}_{ij}$=$c_i c_j^*$ exhibiting purity of 1. Differently from the incoherent case, the subtraction matrix of a coherent single-photon subtraction contains nonzero off-diagonal elements $\chi^{\text{(coh)}}_{ij}\ne0$ for $i\ne j$, which indicates coherence of single-photon subtraction between different modes.

\section{Tomography of a single-photon subtraction}\label{section:tomography}

To experimentally characterize a single-photon subtraction, we employ coherent-state quantum process tomography~\cite{Lobino:2008p6659,Fedorov:2015gg}. As an arbitary quantum state can be expressed in terms of coherent states (the Glauber-Sudarshan P function)~\cite{Sudarshan:1963ts,Glauber:1962tt}, any quantum process can be completely characterized by measuring the responses (the output state and the success probability) on various input coherent states. In general, however, characterizing a multimode process requires a large number of coherent states, which grows exponentially with the number of modes~\cite{Fedorov:2015gg}. For single-photon subtraction, on the other hand, the difficulty of multimode characterization can be circumvented because a coherent state is an eigenstate of any annihilation operator~\cite{Zavatta:2008eq}, i.e., it is not altered by the single-photon subtraction. This fact implies that one can get enough information about single-photon subtraction by measuring only the success probability without measuring the output state. When a coherent state $\ket{\beta}=\ket{\beta b_0}_0 \ket{\beta b_1}_1 \dots \ket{\beta b_{d-1}}_{d-1}$, where $b_{i}$ is the normalized amplitude  ($\sum_i^{d-1} \abs{b_{i}}^2 =1$) for $i$-th mode, and $\abs{\beta}^2$ is the average photon number in the entire modes, is used as an input state of single-photon subtraction in \eq{eq:subtractor}, the output state becomes the same coherent state $\ket{\beta}$, and the success probability is proportional to $\abs{\beta}^2 \left(\sum_{i,j=0}^{d-1} \chi_{ij} b_i b_j^* \right)$. As the success probability depends on the subtraction matrix $\chi$, its element $\chi_{ij}$ can be obtained by measuring the success probabilities for various input coherent states: we can use $\ket{\beta}_i$ and $\ket{\beta}_j$ to interrogate diagonal elements $\chi_{ii}$ and $\chi_{jj}$, and $\ket{\sqrt{\frac{1}{2}} \beta}_i \ket{ \sqrt{\frac{1}{2}} \beta}_j$ and $\ket{ \sqrt{\frac{1}{2}} \beta}_i \ket{\sqrt{-\frac{1}{2}} \beta }_j$ to obtain the real and imaginary values of off-diagonal elements $\chi_{ij}=\chi^*_{ji}$, respectively. In addition, as the subtraction matrix $\chi$ is independent on the input state, it is not necessary to investigate the subtraction matrix as varying the average photon number $\abs{\beta}^2$ of the input coherent states.

\begin{figure}[t]
\centerline{\includegraphics[width=0.475\textwidth]{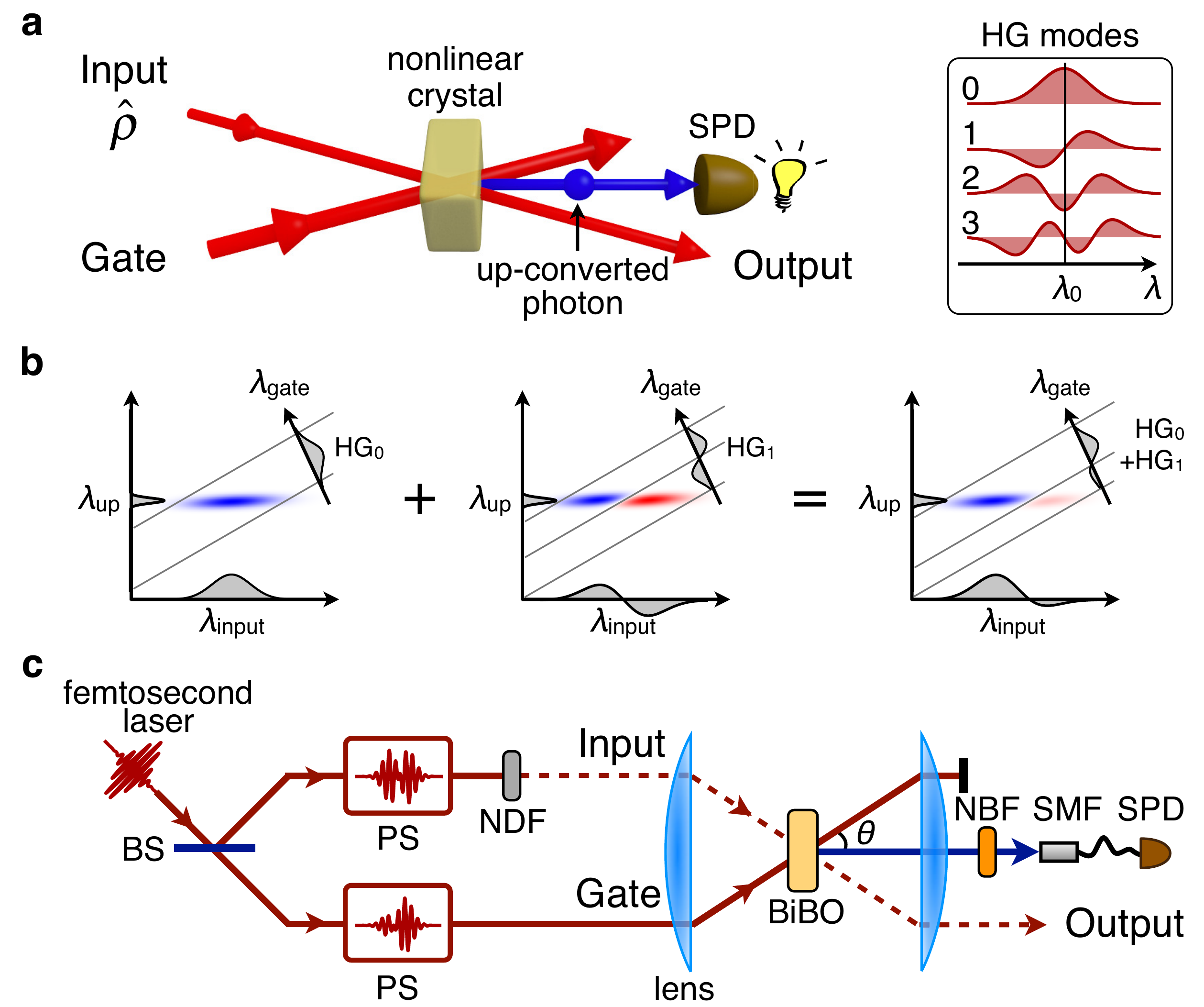}}
\caption{A mode-tunable coherent single-photon subtractor. (a) Conceptual sketch. Detection of the up-converted photon heralds a single-photon subtraction in the input beam, whose time-frequency modes are determined by the spectral amplitude of the strong gate beam. To characterize the single-photon subtraction, weak coherent states are used as the input. Inset: First four Hermite-Gaussian (HG) time-frequency modes, expressed in the wavelength domain. (b) Joint spectral amplitudes of the input and the up-converted beams with HG$_0$ and HG$_1$ gates. The joint spectral amplitudes can be decomposed into the product of the spectral amplitudes of the input and the up-converted beams, drawn as gray filling. As HG$_0$ and HG$_1$ gate beams give rise to the same spectral amplitude for the up-converted beam, sum of the joint spectral amplitudes by HG$_0$ and HG$_1$ gates can also be decomposed into the product of the spectral amplitudes of the input and the up-converted beams. (c) Experimental setup. $\theta=2.5^\circ$; single-photon detector (SPD); non-polarizing beam splitter (BS); pulse shaper (PS); neutral density filter (NDF); narrow bandpass filter (NBF); single-mode fiber (SMF).
}\label{fig:setup}
\end{figure}

\section{Implementation of a mode-tunable coherent single-photon subtractor}\label{section:implementation}

We have implemented a mode-tunable coherent single-photon subtractor for Hermite-Gaussian (HG) time-frequency modes of an input beam based on nonlinear interaction with a strong gate beam, as described in \fig{fig:setup}(a). Inside a second-order nonlinear crystal, photons from the two beams give rise to an up-converted photon via sum frequency generation (SFG). When the up-converted photon is detected by a single-photon detector (SPD),  subtraction of a single photon from the input beam is heralded. In the nonlinear conversion process, the joint spectral amplitude of the input and the up-converted beams is engineered in such a way that the spectral amplitude of the gate beam is directly mapped onto the spectral amplitude of the input beam without affecting the spectral amplitude of the up-converted beam, as shown in~\fig{fig:setup}(b). Such a spectral engineering is accomplished by matching the group velocities of the input and the gate beams~\cite{Eckstein:2011jp,Averchenko:2014dn} and narrow bandpass filtering of the up-converted beam. We can therefore tune the time-frequency modes of the single-photon subtraction by controlling the gate beam: if the gate is in $i$-th HG mode, a single photon is subtracted from the same $i$-th HG mode. In addition, if the gate is in a superposition of different HG modes, a single photon is subtracted coherently from those HG modes because the spectral amplitude of the up-converted beam is independent on the spectral amplitude of the gate beam, see \fig{fig:setup}(b).

Let us assume that the gate beam is a strong coherent state $\ket{\gamma c_0}_0 \ket{\gamma c_1}_1 \dots \ket{\gamma c_{d-1}}_{d-1}$, where the average photon number $\abs{\gamma}^2 \gg1$, and $c_{i}$ is the normalized amplitude ($\sum_i \abs{c_{i}}^2 =1$). Detection of an up-converted photon heralds the coherent single-photon subtraction $\S^{\text{(coh)}}[\hat{\rho}]=\hat{A}_0 \hat{\rho} \hat{A}^{\dagger}_0$ with $\hat{A}_0=\sum_{i=0}^{d-1} (-1)^i c_i \hat{a}_i^{\text{(HG)}}$, where $\hat{a}_i^{\text{(HG)}}$ is the annihilation operator for $i$-th HG mode. The additional coefficient $(-1)^i$ originates from the wavelength inversion with respect to the central wavelength by energy conservation of SFG~\footnote{We consider the case that the frequency bandwidth is much smaller than the central frequency.}, which makes the sign change only for antisymmetric HG modes. In practice, the single-photon subtraction can entail additional annihilation operators $\hat{A}_{n(\ne0)}$ (e.g. due to a nonideal joint spectral amplitude):
\beq
\S^{\text{(SFG)}}[\hat{\rho}] = p_0 \hat{A}_0 \hat{\rho} \hat{A}^\dagger_0 + \sum_{n=1} p_n \hat{A}_n \hat{\rho} \hat{A}_n^\dagger
\eeq
with $\sum_{n=0} p_n=1$. The weight of $\hat{A}_0$, i.e., $p_0$, is defined as \textit{mode selectivity} of single-photon subtraction~\cite{Reddy:2013ip}, which becomes unity for the ideal case. 

\begin{figure*}[t]
\centerline{\includegraphics[width=\textwidth]{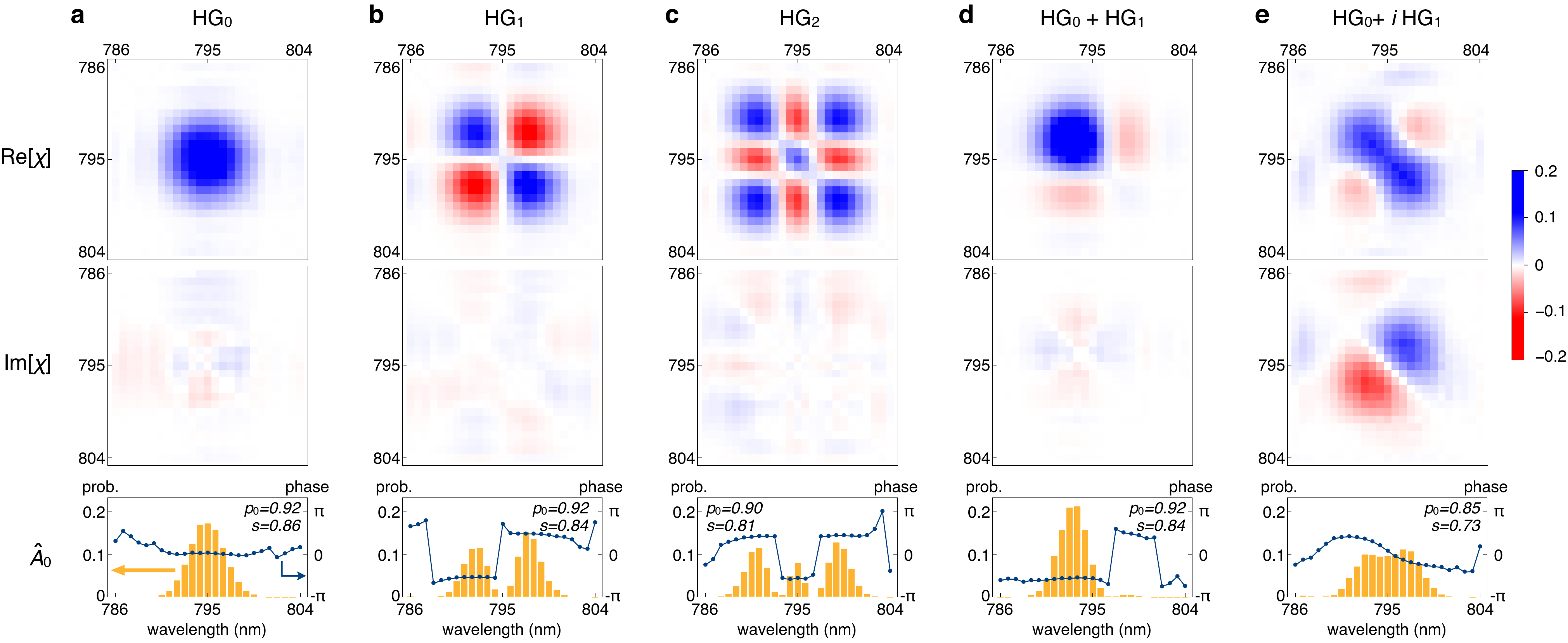}}
\caption{Tomography of single-photon subtraction based on 25 wavelength-band modes. The first, second, and third rows are real and imaginary part of the subtraction matrix $\chi$, and the mode of the dominant annihilation operator $\hat{A}_0$, respectively. In the third row, bars and points represent probability and phase of each wavelength band, respectively, and the line is for visual guide. $p_0$ is mode selectivity, and $s$ is purity.
}\label{fig:pixeldata}
\end{figure*}

Figure \ref{fig:setup}(c) describes the experimental setup developed to implement and characterize the mode-tunable coherent single-photon subtractor. A femtosecond laser (central wavelength: 795 nm, full width at half maximum, FWHM, : 11 nm, repetition rate: 76 MHz) is split into input and gate beams at a beam splitter (BS). The spectral amplitudes of the two beams are individually controlled by pulse shapers (PS) having a spectral resolution of 0.2 nm. 
A neutral density filter (NDF) attenuates the input beam to prepare a coherent state having the average photon number of one per pulse, and the gate beam has 1 mW power (corresponding to around $5 \times10^7$ photons per pulse). The two beams (beam diameter: 1.6 mm) are focused by a single plano-convex lens (focal length: 190 mm) onto a bismuth borate (BiBO) bulk crystal (thickness: 2.5 mm), which generates frequency up-converted light (central wavelength: 397.5nm, FWHM: 0.6 nm) via SFG. To achieve a high mode selectivity, the group velocities of the beams inside the crystal are matched by using the same central wavelength and the same polarization~\cite{Averchenko:2014dn}, and the up-converted light is filtered by a narrow bandpass filter (FWHM: 0.4 nm). The up-converted light is then collected into a single mode fiber, and is detected by an on-off type SPD (Hamamatsu C13001-01, quantum efficiency: 40 \%, dark count rate: 10 Hz). To measure the success probability of single-photon subtraction for the quantum process tomography, we record the count rates of the SPD with various input coherent states.

\section{Experimental results}\label{section:results}

We start with implementing the single-photon subtraction for HG$_0$ mode by sending a gate beam in HG$_0$ mode (central wavelength: 795 nm, FWHM: 4 nm). To represent its subtraction matrix, we choose a wavelength-band mode basis, which consists of 25 different wavelength bands from 786 nm to 804 nm (see Supplementary Information for their spectrums). 
We characterize the implemented subtraction by using input coherent states in the wavelength-band modes, and the average photon number of the input coherent states is increased up to 90 for fast data acquisition.
To construct a physical subtraction matrix (positive and semidefinite), we have employed the maximum likelihood technique~\cite{James:2001p793} for all the following tomography results. Figure \ref{fig:pixeldata}(a) shows the obtained subtraction matrix by using a HG$_0$ gate beam. Note that not only diagonal terms but also off-diagonal terms exist around 795-nm wavelength, manifesting coherent single-photon subtraction from different wavelength-band modes; the imaginary part of the matrix shows negligibly small values because the phase is almost zero over all the wavelength-band modes. 
The dominant eigenvalue of the subtraction matrix, obtained via diagonalization, corresponds to the mode selectivity $p_0$. This being close to one, we can associate the corresponding eigenvector to the dominant single-photon annihilation operator $\hat{A}_0$. It is shown in the last row of \fig{fig:pixeldata}(a), which agrees well with the spectral amplitude of the gate beam and shows a high mode selectivity and purity~\footnote{Note that the purity of the completely mixed subtraction matrix in 25 modes is 0.04.}.

\begin{figure*}[t]
\centerline{\includegraphics[width=0.75\textwidth]{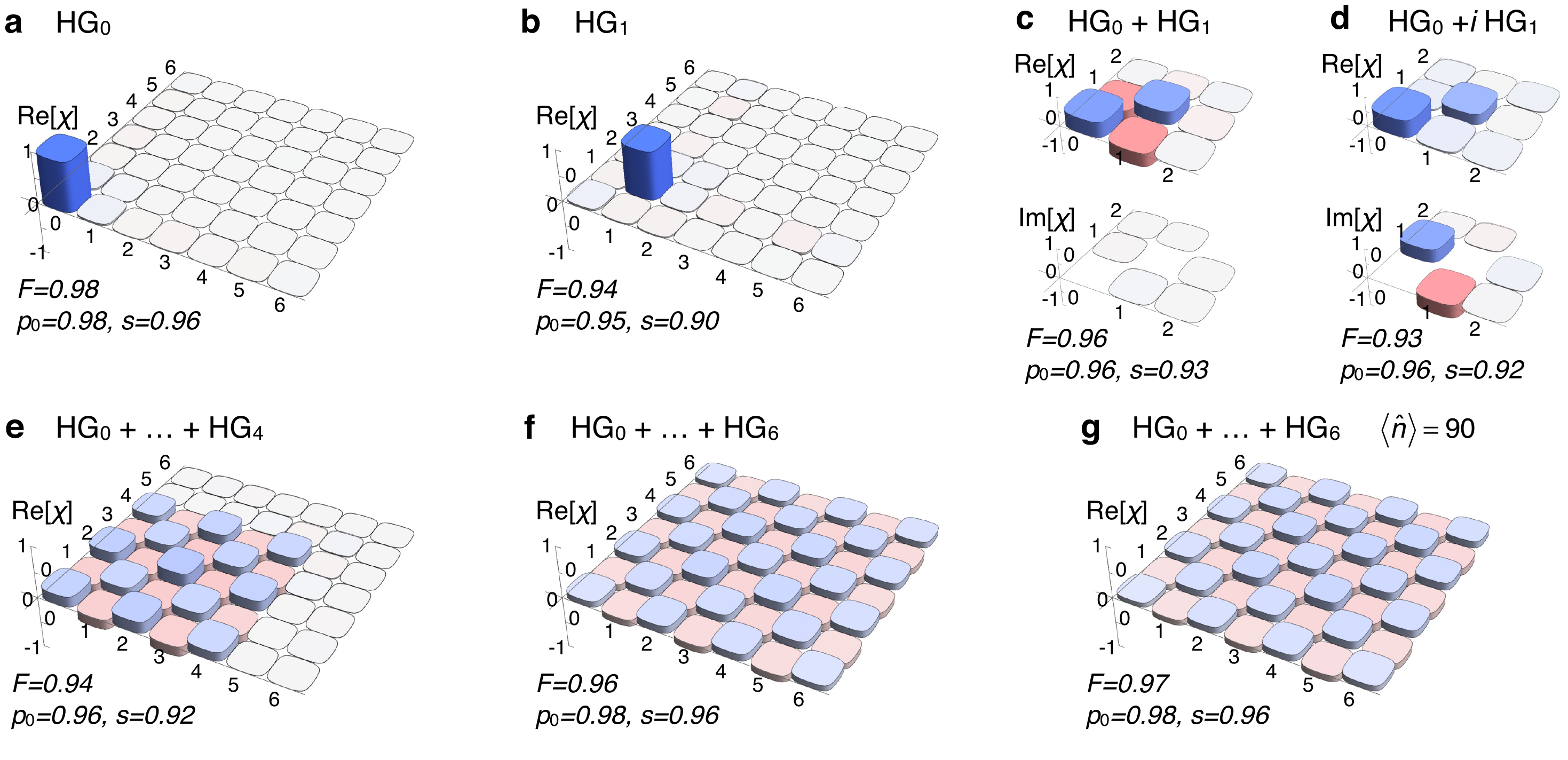}}
\caption{Tomography of single-photon subtraction based on seven Hermite-Gaussian (HG) modes. An index in the horizontal plane denotes the order of a HG mode from 0 to 6. In (a-b) and (e-g), we present only the real part of the subtraction matrix $\chi$ as the imaginary part is negligibly small. For the same reason, we present only a truncated part of a matrix in (c) and (d). Average photon number of probe beam is one in (a-f), and 90 in (g). $F$ is fidelity with the ideal subtraction matrix, $p_0$ is mode selectivity, and $s$ is purity. See Supplementary Information for the full data.
}\label{fig:HGdata}
\end{figure*}

We next tune the single-photon subtractor by adjusting the spectral amplitude of the gate beam. Figure \ref{fig:pixeldata}(b) shows the subtraction matrix obtained by using a HG$_1$ gate beam. The two negative areas in the real part are due to the sign difference of HG$_1$ mode with respect to the central wavelength (see the inset of \fig{fig:setup}(a)); this also confirms the coherence between the longer and the shorter wavelength parts. The sign change also appears as the $\pi$-phase jump in the dominant annihilation operator $\hat{A}_0$, shown in the last row of \fig{fig:pixeldata}(b). Similarly, we implement and characterize single-photon subtraction for HG$_2$ mode, shown in \fig{fig:pixeldata}(c). Figures \ref{fig:pixeldata}(d,e) are obtained by sending a gate beam in a superposition of HG$_0$ and HG$_1$ modes, (d) with $0$-phase difference and (e) with $\pi/2$-phase difference. As HG$_1$ implements $-\hat{a}_1^{\text{(HG)}}$, as discussed in Section~\ref{section:implementation}, the sum of HG$_0$ and HG$_1$ gate modes with $0$-phase difference implements $\frac{1}{\sqrt{2}}{(\hat{a}_0^{\text{(HG)}}-\hat{a}_1^{\text{(HG)}})}$, which makes the subtraction matrix distributed at lower wavelengths than the central wavelength. The $\pi/2$-phase difference between the gate modes results in imaginary values in the subtraction matrix due to the phase difference between wavelength-band modes. We provide additional subtraction matrices by different gate beams in Supplementary Information.

As our single-photon subtractor is designed for parametric multimode sources~\cite{Roslund:2014cb,Lopez:2009be}, we now characterize it with a mode basis approximating the eigenmodes of this process: HG modes \{HG$_0$, HG$_1$, $\dots$, HG$_6$\}.
We have used the same HG modes (central wavelength: 795 nm, FWHM of HG$_0$: 4 nm) for the input beam and the gate beam (see Supplementary Information for the measured spectrum of each HG mode), and have maintained the average photon number of one per pulse for the input beam during the characterization. For a HG$_0$ gate beam, the subtraction matrix, shown in \fig{fig:HGdata}(a), has its dominant element in HG$_0$ mode. It also exhibits high fidelity with the ideal operation $\hat{a}_0^{\text{(HG)}}$ as well as a high mode selectivity and purity. As the gate mode is shifted to higher order, the dominant element in the subtraction matrix is shifted accordingly, see \fig{fig:HGdata}(b) and Supplementary Information. When the gate beam is in a superposition of HG$_0$ and HG$_1$, a coherent single-photon subtraction takes place, as shown in \fig{fig:HGdata}(c) for the same phase and \fig{fig:HGdata}(d) for $\pi/2$-phase difference between the two HG modes. The off-diagonal elements between HG$_0$ and HG$_1$ modes clearly show the coherence between $\hat{a}_0^{\text{(HG)}}$ and $\hat{a}_1^{\text{(HG)}}$ and the tunability of their relative phase. The fidelities with the ideal operations $\frac{1}{\sqrt{2}}(\hat{a}_0^{\text{(HG)}}-\hat{a}_1^{\text{(HG)}})$ and $\frac{1}{\sqrt{2}}(\hat{a}_0^{\text{(HG)}}-i\hat{a}_1^{\text{(HG)}})$, respectively, are also high. The single-photon subtractor can also be tuned to act on multiple HG modes coherently, $\frac{1}{\sqrt{5}}\sum_{i=0}^4 (-1)^i \hat{a}_i^{\text{(HG)}}$ in \fig{fig:HGdata}(e) and $\frac{1}{\sqrt{7}}\sum_{i=0}^6 (-1)^i \hat{a}_i^{\text{(HG)}}$ in \fig{fig:HGdata}(f), respectively. To investigate the independence of the subtraction matrix on the input state, we characterize the single-photon subtraction for \fig{fig:HGdata}(f) using input coherent states with much higher average photon number amounting to 90. The obtained matrix, shown in \fig{fig:HGdata}(g), is almost identical to the subtraction matrix measured by average-photon-number of one in \fig{fig:HGdata}(f), exhibiting fidelity of 0.99 between them. We provide additional subtraction matrices by different gate beams in Supplementary Information.


\section{Discussion}

We discuss here the possible imperfections of the single-photon subtractor by taking into account undesired heralding events. In practice, a single click by an on-off SPD does not always herald single-photon subtraction because it may originate from an accidental event by detector dark counts or detection of two photons~\cite{Kim:2008bw}. A realistic single-photon subtraction $\R$ is then described as
\beq
\R[\hat{\rho}]=w_0\hat{\rho} + w_1 \S[\hat{\rho}] + w_2 \S[\S[\hat{\rho}]],
\label{eq:realsubtractor}
\eeq
which results in the output state $\R[\hat{\rho}]/\tr{(\R[\hat{\rho}])}$ with the success probability proportional to $\tr{(\R[\hat{\rho}])}$. The first term represents the identity operation due to an accidental click, the middle term is the desired single-photon subtraction $\S$ in~\eq{eq:subtractor}, and the last term is the double application of the single-photon subtraction due to two-photon detection. Therefore, their respective weights $w_0$, $w_1$, and $w_2(=1-w_0-w_1)$ are an important factor to assess the quality of the single-photon subtractor. These weights can be measured using input coherent states. If a coherent state $\ket{\beta}_0$ in the dominant subtraction mode is used, the success probability of the operation is proportional to $\tr(\R[\hat{\rho}])=w_0+w_1 p_0 \abs{\beta}^2+w_2 p_0^2 \abs{\beta}^4$; thus, measuring the success probability with respect to $\abs{\beta}^2$ can reveal the weights $w_0$, $w_1$, and $w_2$. Note that the mode selectivity $p_0$ can be obtained through the tomography method presented in Sections~\ref{section:tomography} and~\ref{section:results}. The implemented single-photon subtractor exhibits a dominating contribution of single-photon subtraction ($w_1$=0.99), a very small contribution of the identity operation ($w_0$=0.01), and negligible two-photon subtraction ($w_2<10^{-3}$) (See Supplementary Information for the experimental data). The significant suppression of two-photon subtraction is due to a low conversion ratio ($10^{-3}$) of the input beam to the up-converted beam for a 1-mW gate beam, which still provides a moderate heralding rate of around 2 kHz with an input state of average photon number of one.

\begin{figure}[t]
\centerline{\includegraphics[width=0.4\textwidth]{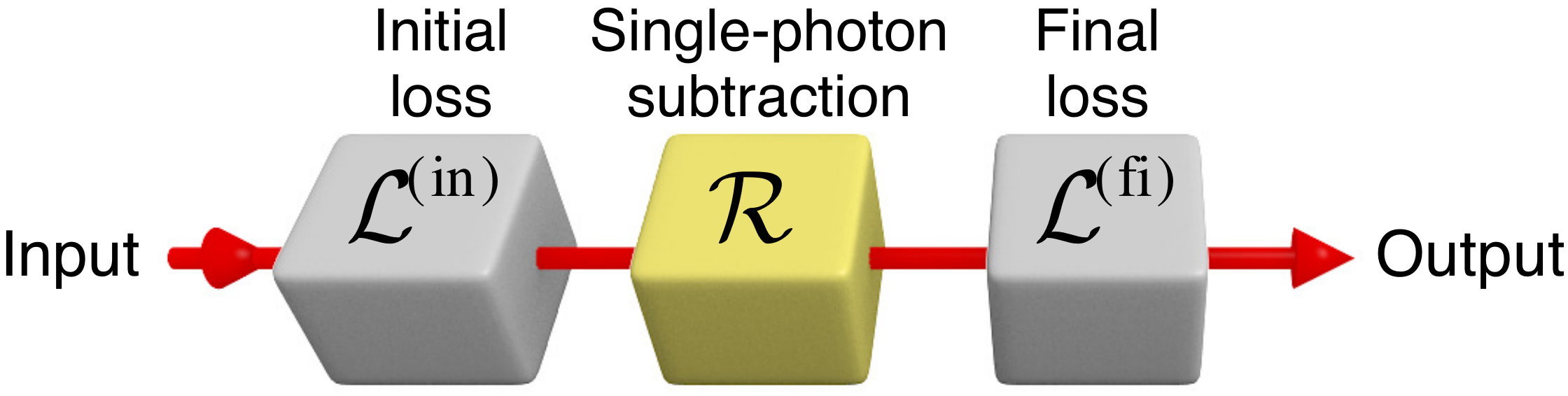}}
\caption{Input state sequentially experiences initial loss $\L^{(\text{in})}$, realistic single-photon subtraction $\R$, and final loss $\L^{(\text{fi})}$.
}\label{fig:loss}
\end{figure}

Based on this realistic model of single-photon subtraction, we can estimate its performance (e.g. negativity of the Wigner function) in a general experimental condition including losses. Figure~\ref{fig:loss} depicts the sequence of operations: initial loss $\L^{(\text{in})}$ accounts for imperfection of quantum state preparation (e.g. excess noise of squeezed vacuum) and the propagation loss before single-photon subtraction, and final loss $\L^{(\text{fi})}$ accounts for the propagation loss after single-photon subtraction and the inefficiency of quantum state measurement (e.g. homodyne detection). Such optical losses can be modeled as a coupling with vacuum by a fictitious beam splitter $\B_T$ (transmittance: $T$). For simplicity, let us consider homogeneous loss for all the modes, $\L^{(\text{in})}=\B_{T^{(\text{in})}}\otimes\B_{T^{(\text{in})}}\otimes\dots$ and $\L^{(\text{fi})}=\B_{T^{(\text{fi})}}\otimes\B_{T^{(\text{fi})}}\otimes\dots$, where $T^{(\text{in})}$ ($T^{(\text{fi})}$) is the transmittance of a fictitious beam splitter for initial (final) loss. If the input state is a multimode state $\hat{\rho}=\hat{\sigma}\otimes\hat{\sigma}\otimes\dots$, which consists of  identically squeezed vacuum $\hat{\sigma}$ in each mode~\cite{Roslund:2014cb}, the final quantum state reduced to the dominant subtraction mode reads
\beq
\hat{\rho}^{\text{(fi)}}_0 &\equiv& \frac{\tr_{12\dots} ( \L^{(\text{fi})}[\R[\L^{(\text{in})}[\hat{\rho}]]] )}  {\tr_{012\dots} ( \L^{(\text{fi})}[\R[\L^{(\text{in})}[\hat{\rho}]]] )} \nonumber \\
&=& r^{\text{(false)}} \B_{T^{(\text{ovr})}}[\hat{\sigma}] +  r^{\text{(corr)}} \frac{\hat{A}_0 \B_{T^{(\text{ovr})}}[\hat{\sigma}] \hat{A}_0^\dagger} {\tr ( \hat{A}_0 \B_{T^{(\text{ovr})}}[\hat{\sigma}] \hat{A}_0^\dagger )},~~~~~
\label{eq:LR}
\eeq
where $T^{(\text{ovr})}(=T^{(\text{fi})}T^{(\text{in})})$ is the overall transmittance of the entire setup, and the two-photon detection weight $w_2$ of $\R$ is set to zero as it is negligible. The first term, $ \B_{T^{(\text{ovr})}}[\hat{\sigma}]$, is a squeezed vacuum mixed with the vacuum noise heralded by a false click in $\R$. It originates from the accidental click (quantified by $w_0=1-w_1$) and single-photon click from other modes (quantified by mode selectivity $p_0$), and has a ratio of $r^{\text{(false)}}=\frac{(1-w_1)+w_1 (1-p_0) T^{(\text{in})} \langle\hat{n}\rangle_{\hat{\sigma}}}  {(1-w_1)+w_1 T^{(\text{in})} \langle\hat{n}\rangle_{\hat{\sigma}}}$, where $\langle\hat{n}\rangle_{\hat{\sigma}}$ is the average photon number of $\hat{\sigma}$. The second term is the single-photon subtracted state from $\B_{T^{(\text{ovr})}}[\hat{\sigma}]$ heralded by a correct click in $\R$, which originates from single-photon subtraction exclusively from the dominant subtraction mode, and has a ratio of $r^{\text{(corr)}}= \frac{w_1 p_0 T^{(\text{in})} \langle\hat{n}\rangle_{\hat{\sigma}}} {(1-w_1)+w_1 T^{(\text{in})} \langle\hat{n}\rangle_{\hat{\sigma}}}$. Based on the characteristics of the implemented single-photon subtractor ($w_1 = 0.99$, $p_0 = 0.9$) and the typical experimental conditions (initial and final losses of 10\%, respectively, which incorporate $2 \%$ optical loss of the implemented single-photon subtractor), one can estimate that a non-Gaussian state exhibiting a negativity of Wigner function amounting to $-\frac{0.3}{2\pi}$ can be obtained from an input state of 4 dB multimode squeezed vacua~\footnote{In the ideal case, a single-photon subtracted squeezed vacuum state or a single-photon Fock state has a negativity of Wigner function amounting to $-\frac{1}{2\pi}$}.


\section{Conclusions}

We have experimentally implemented a mode-tunable coherent single-photon subtractor and characterized it by employing coherent-state quantum process tomography.
We could readily tune the time-frequency modes of single-photon subtraction by adjusting the spectral modes of the gate beam, which does not require a physical reconstruction of a mode-coupling device~\cite{Broome:2013bv,Spring:2013do,Crespi:2013fu,Tillmann:2013jv}. We have implemented various single-photon subtractions such as a subtraction for one HG mode and a coherent subtraction for several HG modes. The subtraction matrices obtained in the wavelength-band modes reveal the modes of the single-photon subtractions in the wavelength domain, and the subtraction matrices in the HG modes directly show the coherence between different HG modes, which is required for parametric multimode sources~\cite{Roslund:2014cb,Lopez:2009be}. A high mode selectivity (typically larger than 0.9) and low imperfections (dark count contribution around 1\% and optical loss around 2 \%) of the single-photon subtractor show its direct applicability to generate multimode non-Gaussian states.

We anticipate that the single-photon subtractor will be an essential operation for a non-Gaussian quantum network, e.g. preparation of hybrid multimode entangled states~\cite{Jeong:2014bl,Morin:2014ip,Andersen:2015dp}, distillation of multipartite entanglement~\cite{Ourjoumtsev:2007he,Kim:2013bt}, measurement-based quantum computing~\cite{Menicucci:2006ir,Ferrini:2013cr}, etc.
In addition, our tomography method is not limited to characterize time-frequency modes, but can be generally applied to other types of light modes such as spatial~\cite{Lopez:2009be}, polarization~\cite{Ra:2016jh}, spatiotemporal~\cite{Gatti:2009je} modes, and it can be extended to characterize a general multimode quantum process~\cite{Fedorov:2015gg}. In particular, it will be useful for identifying couplings among many connected modes (e.g. BosonSampling~\cite{Broome:2013bv,Spring:2013do,Crespi:2013fu,Tillmann:2013jv}, multiple scattering~\cite{Defienne:2016dk}) and more importantly, for quantifying the coherence of such connections~\cite{Tichy:2015jb}.

\begin{acknowledgments}
We thank V. Parigi, J. Roslund, C. Silberhorn, and B. Brecht for fruitful discussions. 
This work is supported by the French National Research Agency projects COMB and SPOCQ, the European Union Grant QCUMbER (no. 665148). C.F. and N.T. are members of the Institut Universitaire de France. Y.-S.R. acknowledges support from the European Commission through Marie Sk\l{}odowska-Curie actions (grant agreement no. 708201).
\end{acknowledgments}



\begin{thebibliography}{10}

\bibitem{Weedbrook:2012fe}
C.~Weedbrook, S.~Pirandola, R.~Garc{\'\i}a-Patr{\'o}n, N.~J. Cerf, T.~C. Ralph,
  J.~H. Shapiro, and S.~Lloyd, Rev. Mod. Phys.{ \bf 84}, 621 (2012).

\bibitem{Yokoyama:2013jp}
S.~Yokoyama, R.~Ukai, S.~C. Armstrong, C.~Sornphiphatphong, T.~Kaji, S.~Suzuki,
  J.~I. Yoshikawa, H.~Yonezawa, N.~C. Menicucci, and A.~Furusawa, Nat. Photonics{ \bf 7}, 982 (2013).

\bibitem{Roslund:2014cb}
J.~Roslund, R.~M. de~Ara{\'u}jo, S.~Jiang, C.~Fabre, and N.~Treps, Nat. Photonics{ \bf 8}, 109 (2014).

\bibitem{Chen:2014jx}
M.~Chen, N.~C. Menicucci, and O.~Pfister, Phys. Rev. Lett.{ \bf 112},
  120505 (2014).

\bibitem{Cai:2016we}
Y.~Cai, J.~Roslund, G.~Ferrini, F.~Arzani, X.~Xu, C.~Fabre, and N.~Treps,
  arXiv.org:1605.02303 (2016).

\bibitem{Broome:2013bv}
M.~A. Broome, A.~Fedrizzi, S.~Rahimi-Keshari, J.~Dove, S.~Aaronson, T.~C.
  Ralph, and A.~G. White, Science (New York, NY){ \bf 339}, 794 (2013).

\bibitem{Spring:2013do}
J.~B. Spring, B.~J. Metcalf, P.~C. Humphreys, W.~S. Kolthammer, X.~M. Jin,
  M.~Barbieri, A.~Datta, N.~Thomas-Peter, N.~K. Langford, D.~Kundys, J.~C.
  Gates, B.~J. Smith, P.~G.~R. Smith, and I.~A. Walmsley, Science (New York,
  NY){ \bf 339}, 798 (2013).

\bibitem{Crespi:2013fu}
A.~Crespi, R.~Osellame, R.~Ramponi, D.~J. Brod, E.~F. Galv{\~a}o, N.~Spagnolo,
  C.~Vitelli, E.~Maiorino, P.~Mataloni, and F.~Sciarrino, Nat. Photonics{ \bf
  7}, 545 (2013).

\bibitem{Tillmann:2013jv}
M.~Tillmann, B.~Dakic, R.~Heilmann, S.~Nolte, A.~Szameit, and P.~Walther,
  Nat. Photonics{ \bf 7}, 540 (2013).

\bibitem{Andersen:2015dp}
U.~L. Andersen, J.~S. Neergaard-Nielsen, P.~Van~Loock, and A.~Furusawa, Nat. Phys.{ \bf 11}, 713 (2015).

\bibitem{Wenger:2004cw}
J.~Wenger, R.~Tualle-Brouri, and P.~Grangier, Phys. Rev. Lett.{ \bf 92},
  153601 (2004).

\bibitem{Lloyd:1999vz}
S.~Lloyd and S.~L. Braunstein, Phys. Rev. Lett.{ \bf 82}, 1784 (1999).

\bibitem{Menicucci:2006ir}
N.~C. Menicucci, P.~van~Loock, M.~Gu, C.~Weedbrook, T.~C. Ralph, and M.~A.
  Nielsen, Phys. Rev. Lett.{ \bf 97}, 110501 (2006).

\bibitem{Bartlett:2002fo}
S.~D. Bartlett, B.~C. Sanders, S.~L. Braunstein, and K.~Nemoto, Phys. Rev.
  Lett.{ \bf 88}, 097904 (2002).

\bibitem{Mari:2012ep}
A.~Mari and J.~Eisert, Phys. Rev. Lett.{ \bf 109}, 230503 (2012).

\bibitem{Ourjoumtsev:2006jn}
A.~Ourjoumtsev, R.~Tualle-Brouri, J.~Laurat, and P.~Grangier, Science (New
  York, NY){ \bf 312}, 83 (2006).

\bibitem{NeergaardNielsen:2006hl}
J. S.~Neergaard-Nielsen, B. M.~Nielsen, C.~Hettich, K.~Molmer, and E. S.~Polzik,
  Phys. Rev. Lett.{ \bf 97}, 083604 (2006).

\bibitem{Ourjoumtsev:2009jh}
A.~Ourjoumtsev, F.~Ferreyrol, R.~Tualle-Brouri, and P.~Grangier, Nat. Phys.{ \bf 5}, 189 (2009).

\bibitem{NeergaardNielsen:2010by}
J.~S. Neergaard-Nielsen, M.~Takeuchi, K.~Wakui, H.~Takahashi, K.~Hayasaka,
  M.~Takeoka, and M.~Sasaki, Phys. Rev. Lett.{ \bf 105}, 053602 (2010).

\bibitem{Jeong:2014bl}
H.~Jeong, A.~Zavatta, M.~Kang, S.~W. Lee, L.~S. Costanzo, S.~Grandi, T.~C.
  Ralph, and M.~Bellini, Nat. Photonics{ \bf 8}, 564 (2014).

\bibitem{Morin:2014ip}
O.~Morin, K.~Huang, J.~Liu, H.~Le~Jeannic, C.~Fabre, and J.~Laurat, Nat. Photonics{ \bf 8}, 570 (2014).

\bibitem{Zavatta:2011ea}
A.~Zavatta, J.~Fiur{\'a}{\v s}ek, and M.~Bellini, Nat. Photonics{ \bf 5}, 52
  (2011).

\bibitem{Ourjoumtsev:2007he}
A.~Ourjoumtsev, A.~Dantan, R.~Tualle-Brouri, and P.~Grangier, Phys. Rev.
  Lett.{ \bf 98}, 030502 (2007).

\bibitem{Takahashi:2010kw}
H.~Takahashi, J.~S. Neergaard-Nielsen, M.~Takeuchi, M.~Takeoka, K.~Hayasaka,
  A.~Furusawa, and M.~Sasaki, Nat. Photonics{ \bf 4}, 178 (2010).

\bibitem{Kim:2008bw}
M.~S. Kim, J. Phys. B: At. Mol. Opt. Phys.{ \bf
  41}, 133001 (2008).

\bibitem{Averchenko:2016gv}
V.~Averchenko, C.~Jacquard, V.~Thiel, C.~Fabre, and N.~Treps, New J. Phys.{ \bf 18}, 083042 (2016).

\bibitem{Kim:2013bt}
H.-J. Kim, J.~Kim, and H.~Nha, Phys. Rev. A{ \bf 88}, 032109 (2013).

\bibitem{Eckstein:2011jp}
A.~Eckstein, B.~Brecht, and C.~Silberhorn, Opt. Express{ \bf 19}, 13770
  (2011).

\bibitem{Brecht:2014eg}
B.~Brecht, A.~Eckstein, R.~Ricken, V.~Quiring, H.~Suche, L.~Sansoni, and
  C.~Silberhorn, Phys. Rev. A{ \bf 90}, 030302 (2014).

\bibitem{Averchenko:2014dn}
V.~A. Averchenko, V.~Thiel, and N.~Treps, Phys. Rev. A{ \bf 89}, 063808
  (2014).

\bibitem{Manurkar:2016hx}
P.~Manurkar, N.~Jain, M.~Silver, Y.-P. Huang, C.~Langrock, M.~M. Fejer,
  P.~Kumar, and G.~S. Kanter, Optica{ \bf 3}, 1300 (2016).

\bibitem{Lobino:2008p6659}
M.~Lobino, D.~Korystov, C.~Kupchak, E.~Figueroa, B.~C. Sanders, and A.~I.
  Lvovsky, Science (New York, NY){ \bf 322}, 563 (2008).

\bibitem{Fedorov:2015gg}
I.~A. Fedorov, A.~K. Fedorov, Y.~V. Kurochkin, and A.~I. Lvovsky, New J. Phys.{ \bf 17}, 043063 (2015).

\bibitem{Kumar:2013ic}
R.~Kumar, E.~Barrios, C.~Kupchak, and A.~I. Lvovsky, Phys. Rev. Lett.{
  \bf 110}, 130403 (2013).

\bibitem{Sudarshan:1963ts}
E.~C.~G. Sudarshan, Phys. Rev. Lett.{ \bf 10}, 277 (1963).

\bibitem{Glauber:1962tt}
R.~J. Glauber, Phys. Rev. Lett.{ \bf 10}, 84 (1963).

\bibitem{Zavatta:2008eq}
A.~Zavatta, V.~Parigi, M.~S. Kim, and M.~Bellini, New J. Phys.{ \bf
  10}, 123006 (2008).

\bibitem{Note1}
We consider the case that the frequency bandwidth is much smaller than the central frequency.

\bibitem{Reddy:2013ip}
D.~V. Reddy, M.~G. Raymer, C.~J. McKinstrie, L.~Mejling, and K.~Rottwitt,
  Opt. Express{ \bf 21}, 13840 (2013).

\bibitem{James:2001p793}
D. F. V.~James, P.~G. Kwiat, W.~J. Munro, and A. G.~White, Phys. Rev. A{ \bf 64},
  052312 (2001).

\bibitem{Note2}
Note that the purity of the completely mixed subtraction matrix in 25 modes is
  0.04.

\bibitem{Lopez:2009be}
L.~Lopez, B.~Chalopin, A.~R. de~la Souch{\`e}re, C.~Fabre, A.~Ma{\^\i}tre, and
  N.~Treps, Phys. Rev. A{ \bf 80}, 043816 (2009).

\bibitem{Note3}
In the ideal case, a single-photon subtracted squeezed vacuum state or a
  single-photon Fock state has a negativity of Wigner function amounting to
  $-\protect \frac {1}{2\pi }$.

\bibitem{Ferrini:2013cr}
G.~Ferrini, J.~P. Gazeau, T.~Coudreau, C.~Fabre, and N.~Treps, New J. Phys.{ \bf 15}, 093015 (2013).

\bibitem{Ra:2016jh}
Y.-S. Ra, H.-T. Lim, and Y.-H. Kim, Phys. Rev. A{ \bf 94}, 042329 (2016).

\bibitem{Gatti:2009je}
A.~Gatti, E.~Brambilla, L.~Caspani, O.~Jedrkiewicz, and L.~A. Lugiato, Phys.
  Rev. Lett.{ \bf 102}, 223601 (2009).

\bibitem{Defienne:2016dk}
H.~Defienne, M.~Barbieri, I.~A. Walmsley, B.~J. Smith, and S.~Gigan, Science
  Advances{ \bf 2}, e1501054 (2016).

\bibitem{Tichy:2015jb}
M.~C. Tichy, Y.-S. Ra, H.-T. Lim, C.~Gneiting, Y.-H. Kim, and K.~M{\o}lmer, New J. Phys.{ \bf 17}, 023008 (2015).

\end{thebibliography}

\end{document}